# Practical Implementation of Link Adaptation with Dual Polarized Modulation

Anxo Tato, Carlos Mosquera
*atlanTTic Research Center, University of Vigo*
Vigo, Spain
anxotato@gts.uvigo.es, mosquera@gts.uvigo.es

Pol Henarejos, Ana Pérez-Neira
*Centre Tecnològic de Telecomunicacions de Catalunya (CTTC)*
Castelldefels, Spain
pol.henarejos@cttc.es, ana.perez@cttc.es

*Abstract*—The use of dual polarization in mobile satellite systems is very promising for increasing the channel capacity. Polarized Modulation is proposed in this paper for use in practical systems, by providing simple equations for computing its capacity and featuring a link adaptation algorithm. This scheme shows remarkable gains in the spectral efficiency when compared with single polarization and other multi-antenna techniques such as V-BLAST. Polarized Modulation is a particular instance of more general Index Modulations, which are being considered for 5G networks. Thus, the proposed link adaptation algorithm could find synergies with current activities for future terrestrial networks.

## I. INTRODUCTION

The simultaneous use of two polarizations is a means to increase the spectral efficiency that is being explored for mobile satellite communications [1]. At low frequencies, like L- and S-band, only one polarization has been traditionally used due to fear to small Cross-Polar-Discriminations (XPD) [2]. Although Right Hand Circular Polarization (RHCP) was the typical option, it is possible to use simultaneously two orthogonal polarizations, RHCP and Left Hand Circular Polarization (LHCP), to communicate with users. This is analogous to a 2x2 Multiple-Input-Multiple-Output (MIMO) system, simply replacing the spatial by the polarization component. Therefore, the application of MIMO signal processing techniques makes it possible to achieve throughput gains while maintaining the same transmit power, even in the presence of interference between polarizations.

There are different MIMO modes which can be employed in the scenario of a mobile satellite system with Dual Polarization (DP) but, in practice, only those not requiring Channel State Information at the Transmitter (CSIT) are of interest because of the long propagation delay. Two candidates are Orthogonal Polarization-Time Block Code (OPTBC), based on 2x2 Alamouti Space-Time Coding, and Vertical-Bell Laboratory Layered Space-Time (V-BLAST), a spatial multiplexing technique which transmits two independent streams of symbols, one per each polarization.

This work is funded by projects MYRADA (TEC2016-75103-C2-2-R) and ELISA (TEC2014-59255-C3-1-R).

The DP mobile satellite system proposed in [1] and [3] allows the transmitter to switch among different MIMO modes across frames to maximize the spectral efficiency. Similar adaptation mechanisms take place in other systems such as Long Term Evolution (LTE) [4]. In addition to OPTBC and V-BLAST, in [3] an additional MIMO mode named Polarized Modulation (PMod) is proposed since it provides many advantages. PMod is a particular type of an Index Modulation which allows a remarkable gain in spectral efficiency at intermediate SNRs (Signal to Noise Ratios) which cannot be achieved by just employing OPTBC and V-BLAST.

In PMod two streams of bits are conveyed, one that encodes which symbols of the constellation are transmitted and another stream that indicates which one of the two polarizations is employed to radiate each symbol. Previous works that deal with link adaptation for Index Modulations, like [5] does with Spatial Modulation, only consider adaptive modulation, i.e., the selection of the modulation order. The novelty of this paper is that we propose the use of two independent variable rate channel encoders, one for each bit stream. In this way the coding rate of each encoder is adapted independently according to the instantaneous channel capacity of each component, the polarizations and the symbols.

The aim of this paper is the description of a link adaptation procedure for selecting the Modulation and Coding Scheme (MCS) to be used in a practical implementation of DP system with PMod MIMO mode. Our idea makes use of two independent Lookup Tables (LUT) for selecting the coding rates of each bit stream. Some adaptive margins are added to the SNRs at the input of each LUT with the objective to guarantee a predefined Frame Error Rate (FER). Moreover, simulation results are included to show the benefits of exploiting PMod among the MIMO modes of a DP system. Lastly, some results of the capacity for PMod based on recent findings [6] are introduced since they are required in the link adaptation process.

The remainder of the paper is structured as follows. Section II provides a brief description of the DP satellite communication system. Section III deals with the capacity computation of the PMod scheme. Then, Section IV describes the Physical Layer Abstraction techniques used in the simulations. After that,

Section V introduces the link adaptation algorithm for MCS selection and Section VI provides some simulation results. Lastly, the main conclusions are collected.

## II. SYSTEM MODEL

The forward link of a satellite communication system which serves mobile users in the L-band is considered here, with a transmitter (the gateway) communicating with a receiver (the Mobile Terminal). Both are equipped with DP antennas and the transmitter always uses the PMod scheme, a particular case of Index Modulations [7] which exploits the polarization flexibility.

PMod combines two sources for conveying information: the symbol $s$, taken from a given constellation $\mathcal{S}$, and the polarization used to transmit that symbol, $l$. Hence, in each frame two streams of bits are transmitted: one encoding the radiated symbols and a second stream signalling the polarization of each symbol.

Only link adaptation in the forward link is considered here, therefore, the transmitter has to decide the physical layer parameters for each transmitted frame. Fig. 1 shows the architecture of the PMod transmitter with Adaptive Coding and Modulation (ACM), with two parallel channel encoders with variable and independent coding rates ($r_S$ and $r_P$) applied to each stream of bits (frames of $N$ symbols are assumed). The purpose of the link adaptation algorithm is to select the MCS, i.e. the coding rate, used in each channel encoder: $m_{Si}$ for the bits which encode the symbols and $m_{Pi}$ for the bits which encode the polarizations.

We also assume that there exists a return channel used by the receiver to feedback the information the link adaptation algorithm needs at the transmitter side: the two effective SNRs and the outcome of the decodification of each codeword, one for each part, symbols and polarizations.

The system model of a general DP system for a given discrete time instant is

$$\mathbf{y} = \sqrt{\gamma}\mathbf{H}\mathbf{x} + \mathbf{w}, \quad (1)$$

where $\mathbf{y} \in \mathbb{C}^2$ is the received vector, $\gamma$ is the average SNR, $\mathbf{H} = (\mathbf{h}_1\ \mathbf{h}_2) \in \mathbb{C}^{2\times 2}$ is the channel matrix, $\mathbf{x} \in \mathbb{C}^2$ is the transmitted signal and $\mathbf{w} \sim \mathcal{CN}(\mathbf{0}, \mathbf{I}_2)$ is the Additive White Gaussian Noise (AWGN). With PMod, since $\mathbf{x}$ has always a zero component, (1) can be expressed as $\mathbf{y} = \sqrt{\gamma}\mathbf{h}_l s + \mathbf{w}$, with $\mathbf{h}_l$ the $l$-column of the channel matrix, where the index $l$ selects the polarization that transmits the complex symbol $s$.

## III. PMOD CAPACITY

In PMod, since the symbol $s$ and the hopping index which selects the polarizations $l$ transmit information, the Mutual Information (MI) between the inputs and the output of the channel can be expressed as

$$I(\mathbf{y}; s, l) = I(\mathbf{y}; s|l) + I(\mathbf{y}; l) = I_S + I_P, \quad (2)$$

where the MI is expressed as the sum of two components: the capacity of the symbols $I_S$ and the polarization bit capacity $I_P$. The capacity of IM is studied in [6] and [8], among others. In [6] a first order approximation of (2) for a general Index Modulation is obtained. From the equation obtained there, which is actually an upper bound, we derive the following approximation of the term $I_P$ for the particular case of PMod:

$$I_P \simeq \log_2\left(\frac{2}{1 + e^{-\gamma|s|^2\|\mathbf{h}_1 - \mathbf{h}_2\|^2}}\right). \quad (3)$$

The capacity $I_P$ depends on the average SNR $\gamma$, the energy of the symbols and the distance between the two columns of the channel matrix $\mathbf{H}$. Equation (3) will be used later to calculate the instantaneous polarization bit capacity in (9).

For an AWGN channel, with a normalized diagonal channel matrix $\mathbf{H} = \sqrt{2}\mathbf{I}_2$, and for symbols $s$ with unit energy, (3) is transformed into this equation,

$$I_P \simeq \log_2\left(\frac{2}{1 + e^{-4\gamma}}\right), \quad (4)$$

which gives an upper bound of the polarization bit capacity for a diagonal channel. This capacity is represented with a black line in Fig. 2.

However, since the polarization bit capacity $I_P$ with that diagonal channel matrix is used always in the last step of the effective SNR calculation, a more accurate equation than (4) will be derived. The real $I_P$, which can be obtained using Monte Carlo to approximate the expectation of Equation (6) of [6], is plotted with a blue line in Fig. 2. Following [9] we have also found that an exponential function of the form $I_P \simeq 1 - e^{-\alpha\gamma}$ can perform even better than (4). The value of $\alpha$ estimated using curve fitting with the nonlinear least squares method turns out to be 1.30. Fig. 2 shows how this approximation matches tightly the Monte Carlo values. Hereafter this capacity will be referred as $\Phi_P(\gamma)$:

$$\Phi_P(\gamma) = 1 - e^{-1.30\gamma}. \quad (5)$$

For the sake of completeness, we also include in Fig. 2 the constrained capacity of a Quadrature Phase Shift Keying (QPSK) constellation in an AWGN channel, $\Phi_S(\gamma)$, since we will refer to this function in the next section. This capacity can be also approximated by sums of exponential functions as proposed in [9]. Again, we employ the same curve fitting procedure to obtain an analytical expression which matches the QPSK capacity calculated with Monte Carlo. Finally, the expression obtained for the QPSK capacity is

$$\Phi_S(\gamma) = 2\cdot\left(1 - 0.8551 \cdot e^{-0.5718\gamma} - (1 - 0.8551) \cdot e^{-1.55\gamma}\right). \quad (6)$$

## IV. PHYSICAL LAYER ABSTRACTION IN PMOD

In order to avoid long simulations, resorting to Physical Layer Abstraction techniques [10] is a common practice. A metric named effective SNR condenses in a single value, SNR$_{\text{eff}}$, all the channel variations during a frame spanning $N$ symbols. By referring the effective SNR to the thresholds of the different MCS, the success of the communication can be determined.

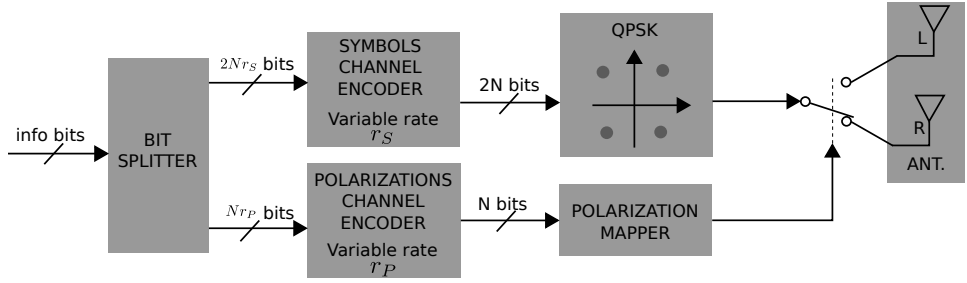

Fig. 1. Architecture of the PMod transmitter with ACM.

TABLE I
SET OF MCS FOR THE SYMBOLS (QPSK CONSTELLATION IS ASSUMED).

| Coding rate | 0.34 | 0.40 | 0.48 | 0.55 | 0.63 | 0.70 | 0.77 | 0.83 | 0.87 |
|---|---|---|---|---|---|---|---|---|---|
| Spectral efficiency | 0.68 | 0.80 | 0.96 | 1.10 | 1.26 | 1.40 | 1.54 | 1.66 | 1.74 |
| Threshold SNR ($\text{SNR}_{th}$) (dB) | -2.15 | -1.21 | -0.09 | 0.83 | 1.84 | 2.74 | 3.67 | 4.54 | 5.19 |

TABLE II
SET OF MCS FOR THE POLARIZATIONS.

| Coding rate | 0.1 | 0.2 | 0.3 | 0.4 | 0.5 | 0.6 | 0.7 | 0.8 | 0.9 |
|---|---|---|---|---|---|---|---|---|---|
| Threshold SNR ($\text{SNR}_{th}$) (dB) | -10.91 | -7.65 | -5.62 | -4.06 | -2.73 | -1.52 | -0.33 | 0.93 | 2.48 |

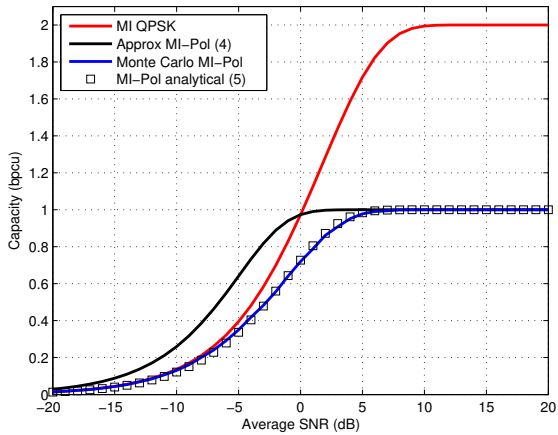

Fig. 2. Mutual Information for symbol and polarization in PMod.

Since independent encoding is applied to the source bits conveyed by the symbols and the polarization choice, two effective SNR values have to be computed. The calculation of the effective SNR of the symbols requires the computation of the SNR of each received symbol $\gamma_n$, which for a Maximum Ratio Combining (MRC) Single-Input-Multiple-Output (SIMO) receiver [11] is expressed as

$$\gamma_{n,k} = \gamma \frac{|[\mathbf{H}_n]_{1k}|^4 + |[\mathbf{H}_n]_{2k}|^4 + 2|[\mathbf{H}_n]_{1k}|^2|[\mathbf{H}_n]_{2k}|^2}{|[\mathbf{H}_n]_{1k}|^2 + |[\mathbf{H}_n]_{2k}|^2}. \quad (7)$$

In the previous equation, $[\mathbf{H}_n]_{ij}$ denotes the coefficient $(i,j)$ of the channel matrix $\mathbf{H}_n$ for time instant $n$ and $k = \{1, 2\}$ refers to the index of the polarization. There are two values $\gamma_{n,k}$, one for each polarization, which are used to compute their respective effective SNR. The final effective SNR is $\text{SNR}_{\text{eff,S}} = \min(\text{SNR}_{\text{eff,S,1}}, \text{SNR}_{\text{eff,S,2}})$. The effective SNR of the symbols of polarization $k$ is obtained as

$$\text{SNR}_{\text{eff,S,k}} = \Phi_S^{-1}\left(\frac{1}{N}\sum_{n=1}^{N}\Phi_S(\gamma_{n,k})\right). \quad (8)$$

The effective SNR of the bits encoding the polarization $\text{SNR}_{\text{eff,P}}$ can be obtained as

$$\text{SNR}_{\text{eff,P}} = \Phi_P^{-1}\left(\frac{1}{N}\sum_{n=1}^{N} I_P(\mathbf{H}_n)\right) \quad (9)$$

where $I_P(\mathbf{H}_n)$ is computed as in (3). Even though (3) is an upper bound for the transmission capacity of the polarization selection, this is not a problem in practice. The use of an adaptive margin in the link adaptation algorithm, serves to compensate for the potential errors in the capacity evaluation due to, among other factors, the outdated CSIT. This approximation is used to avoid computing the capacity at each time instant with Monte Carlo, which would be prohibitive.

To sum up, each frame carries two codewords, one in the symbols and other in the polarization hops. The outcome of the decodification of each codeword is predicted using their respective effective SNR, $\text{SNR}_{\text{eff,S}}$ or $\text{SNR}_{\text{eff,P}}$. If the effective SNR of a codeword is higher than the threshold SNR of the MCS used to transmit that codeword, this is considered to be correctly decoded and if it is lower, an error is assumed to occur in the decodification.

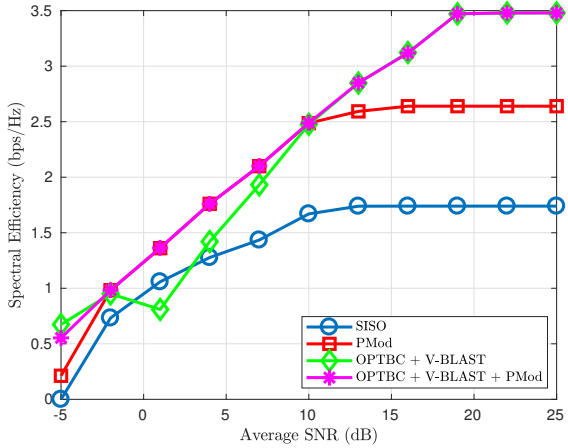

Fig. 3. Spectral efficiency of the simulations.

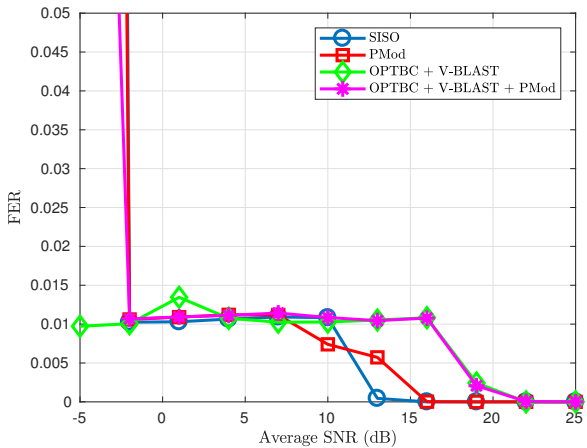

Fig. 4. Frame Error Rate of the simulations ($p_0 = 0.01$).

## V. Algorithm for MCS selection

The selection of both MCS is done by means of two Lookup Tables (LUTs), represented by two step functions, $\Pi_S(\cdot)$ and $\Pi_P(\cdot)$, which map SNR intervals to MCS following Tables I-II. On the one hand, the threshold SNRs of Table I are obtained using the function $\Phi_S(\gamma)$ which gives the capacity of a QPSK modulation in an AWGN channel. On the other hand, the threshold SNRs for polarization MCS shown in Table II are taken for a diagonal channel matrix and AWGN noise and their values are calculated using the function $\Phi_P(\gamma)$.

The equations to select the MCS are

$$m_{Si} = \Pi_S(\text{SNR}_{Si-d} + c_{Si}), \quad m_{Pi} = \Pi_P(\text{SNR}_{Pi-d} + c_{Pi}), \tag{10}$$

where $\text{SNR}_{Si-d}$ and $\text{SNR}_{Pi-d}$ are the effective SNR for the symbols and the polarizations respectively, calculated by the receiver and sent back to the transmitter through a feedback channel − a Round Trip Time (RTT) of $d$ frames is assumed. $c_{Si}$ and $c_{Pi}$ are two adaptive margins updated by the transmitter by using the corresponding received feedback from the other end. The receiver also informs the transmitter if the codewords of the symbols and the polarizations were decoded correctly or not by sending two binary variables, $\epsilon_{Si-d}$ and $\epsilon_{Pi-d}$, which take the value 1 if an error (NAK) has occurred and 0 otherwise (ACK).

As in [3], margins are updated by a recursive equation obtained from the optimization problem $\min_c |\mathbb{E}[\epsilon] - p_0/2|^2$, which is based on finding a margin that adjusts the average number of errors in each stream to a predefined value $p_0/2$. Hereafter, only those frames with both codewords (for symbols and polarizations) decoded successfully count as correct for the Frame Error Rate (FER) estimation. Therefore, the margins obtained here are aimed to guarantee a global FER of $p_0$.

The two equations for updating the margins are

$$c_{S,i+1} = c_{S,i} - \mu(\epsilon_{S,i-d} - p_0/2) \tag{11}$$

and

$$c_{P,i+1} = c_{P,i} - \mu(\epsilon_{P,i-d} - p_0/2), \tag{12}$$

where $\mu$ is an adaptation step which in our simulations take the value 0.05. Note that there are two independent margins, $c_{Si}$ and $c_{Pi}$, each one being updated by using their respective acknowledgement ($\epsilon_{S,i-d}$ and $\epsilon_{P,i-d}$). When an ACK has been received the margin is increased by a small quantity, $\mu \cdot p_0/2$, and when an error is reported ($\epsilon_{i-d} = 1$) the margin is reduced more drastically by a quantity $\mu \cdot (1 - p_0/2)$, which for small values of the objective FER is approximately $\mu$.

## VI. Simulation results

All the simulation results presented here are obtained for a maritime scenario with a vessel moving at a constant speed of 50 km/h. The parameters of the channel generator of the DP mobile satellite system are taken from [12]. The carrier frequency in the simulations is 1.6 GHz. Frames of 80 ms and 2560 symbols are used in the physical layer, similarly to the bearer F80T1Q-1B of ETSI TS 102-744 [13]. Each simulation comprises the transmission and reception of $M = 40,000$ frames for a specific average SNR. Average spectral efficiency, defined as $1/M \sum_{i=1}^{M}(1 - (\epsilon_{Si} + \epsilon_{Pi}))(r_{m_{Si}} + r_{m_{Pi}})$, and cumulative FER during the whole transmission is also computed for each simulation. In the previous expression $r_{m_{Si}}$ and $r_{m_{Pi}}$ are the rates of the MCS selected for frame $i$ and ($\epsilon_{Si} + \epsilon_{Pi}$) is the logical OR operation between the two acknowledgements. Lastly, the RTT is set to 7 frames (560 ms) to emulate the feedback delay for geostationary (GEO) satellites.

Fig. 3 plots the average spectral efficiency for four sets of simulations over a range of average SNRs from -5 to 25 dB. ACM is used in all cases, with the link adaptation algorithm described in this paper for the simulations with PMod and with a similar algorithm for the other modes, as described in [3]. For those simulations with several available MIMO modes, the mode can be switched from frame to frame and its selection is performed according to [3], selecting the mode that maximizes

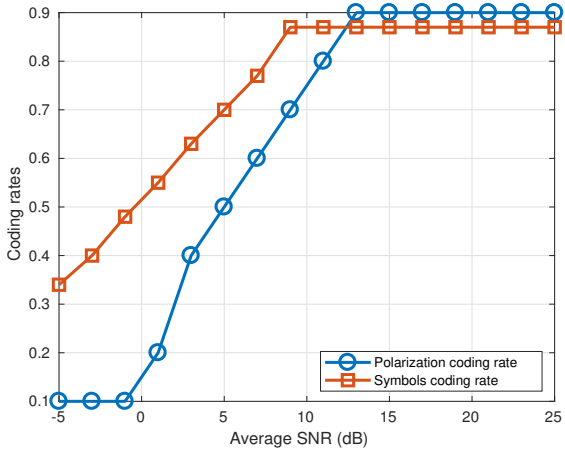

Fig. 5. Most used coding rate for the symbols channel encoder and polarizations channel encoder.

the spectral efficiency given the current value of the effective SNRs of all modes.

Firstly, the comparison between mono polarization (SISO curve) and DP in Fig. 3 shows that an increment of up to +50 % in the spectral efficiency can be achieved with PMod as expected when a QPSK constellation is used. For low SNRs, this realistic evaluation of PMod shows that in this scenario there is always a minimum gain of +30 %. These gains can be employed to serve more users or to increase the data rate of the individual users.

Note that the same constellation −QPSK− has been used for both SISO and MIMO configurations. In order to match the performance of DP-PMod, the SISO link would require higher order constellations. With 8-PSK, the SISO link would perform as DP-PMod with QPSK, only for SNR high enough. More in general, the channel capacity is higher for DP-PMod than for SISO, as easily inferred from Equation (2).

Secondly, SISO and PMod are compared with other two DP systems where several MIMO modes are available. In some simulations the modes are only OPTBC and V-BLAST, whereas in other cases these two modes are complemented with PMod. From Fig. 3 it is clear that the inclusion of PMod among the MIMO modes of the DP system is worthwhile. For a range of SNRs from −2 to 7 dB, PMod provides a spectral efficiency gain with respect to a configuration where only OPTBC and V-BLAST are used. Thirdly, for systems where high SNRs are achievable, the V-BLAST mode allows to double the spectral efficiency of the Single Polarization system. Finally, Fig. 4 confirms that the use of the adaptive margin guarantees a FER around the objective value ($p_0 = 0.01$ in this case) for a wide range of SNRs and for all the MIMO modes considered here.

Lastly, in Fig. 5 we show the most used coding rate for the symbols channel encoder and for the polarizations channel encoder as function of the average SNR of the channel $\gamma$. As expected, for lower SNRs more protected MCS are selected, i.e., with lower coding rates, and for higher SNRs more efficient MCS are chosen. If we compare this figure with Tables I-II we observe that the effective SNR for the polarization $\text{SNR}_{\text{eff,P}}$ is lower than the average SNR of the channel $\gamma$ since the channel matrices of the simulated scenario are no longer diagonal as Table II assumes. This leads to the selection of lower rates than one could expect from looking at $\gamma$. And, on the other hand, since MRC is used at the receiver, the power received at each polarization contributes to the final SNR of the symbols, leading to the selection of higher coding rates for the symbols that one could expect from looking the $\gamma$. All this shows the adequacy of using two independent variable rate channel encoders, one for the string of bits which selects the constellation symbols and another for the string of bits which selects the polarizations hops.

## VII. Conclusions

In this paper we have introduced a novel architecture for a Polarized Modulation transmitter based on two independent channel encoders, one for bits encoding the symbols and another for bits encoding the polarizations. Analytical expressions to estimate the capacity of each PMOd component are provided, enabling the use of an optimized coding rate in each channel encoder to match their respective channel capacities. A link adaptation algorithm based on a Lookup Table per each component with two independent adaptive margins is developed. Simulation results show how PMod outperforms other MIMO modes for an important range of operating SNRs and also how the use of an adaptive margin guarantees a fixed Frame Error Rate. Lastly, the dependence of the selected coding rates with the SNR shows the suitability of using two independent channel encoders.

## References

[1] P. Henarejos, A. Perez-Neira, N. Mazzali, and C. Mosquera, "Advanced signal processing techniques for fixed and mobile satellite communications," in *2016 8th Advanced Satellite Multimedia Systems Conference and the 14th Signal Processing for Space Communications Workshop (ASMS/SPSC)*, Spain, Mallorca, Sept 2016, pp. 1–8.

[2] M. Richharia, *Mobile Satellite Communications: Principles and Trends*, 2nd ed. Wiley, 2014.

[3] A. Tato, P. Henarejos, C. Mosquera, and A. Pérez-Neira, "Link Adaptation Algorithms for Dual Polarization Mobile Satellite Systems," in *Wireless and Satellite Systems*, P. Pillai, K. Sithamparanathan, G. Giambene, M. Á. Vázquez, and P. D. Mitchell, Eds. Cham: Springer International Publishing, 2018, pp. 52–61.

[4] Q. Li, G. Li, W. Lee, M. i. Lee, D. Mazzarese, B. Clerckx, and Z. Li, "MIMO techniques in WiMAX and LTE: a feature overview," *IEEE Communications Magazine*, vol. 48, no. 5, pp. 86–92, May 2010.

[5] P. Yang, M. D. Renzo, Y. Xiao, S. Li, and L. Hanzo, "Design guidelines for spatial modulation," *IEEE Communications Surveys Tutorials*, vol. 17, no. 1, pp. 6–26, Firstquarter 2015.

[6] P. Henarejos, A. Perez-Neira, A. Tato, and C. Mosquera, "Channel dependent Mutal Information in Index Modulations," in *Proceedings. (ICASSP '18). IEEE International Conference on Acoustics, Speech, and Signal Processing, 2018.* Available at http://gpsc.uvigo.es/sites/default/files/publications/ICASSP-CTTC-Pol-vf.pdf, 2018, pp. 1–15.

[7] E. Basar, "Index modulation techniques for 5G wireless networks," *IEEE Communications Magazine*, vol. 54, no. 7, pp. 168–175, July 2016.

[8] P. Henarejos and A. I. Perez-Neira, "Capacity Analysis of Index Modulations Over Spatial, Polarization, and Frequency Dimensions," *IEEE Transactions on Communications*, vol. 65, no. 12, pp. 5280–5292, Dec 2017.

[9] J. Arnau, A. Rico-Alvarino, and C. Mosquera, "Adaptive transmission techniques for mobile satellite links," in *Proc. AIAA ICSSC*, Ottawa, Canada, Sep. 2012.

[10] F. Kaltenberger, I. Latif, and R. Knopp, "On scalability, robustness and accuracy of physical layer abstraction for large-scale system-level evaluations of LTE networks," in *2013 Asilomar Conference on Signals, Systems and Computers*, Pacific Grove, California, Nov 2013, pp. 1644–1648.

[11] P. Henarejos, "Polarization and Index Modulations: a Theoretical and Practical Perspective," Ph.D. dissertation, Universitat Politécnica de Catalunya, Departament de Teoria del Senyal i Comunicacions, 2017.

[12] M. Sellathurai, P. Guinand, and J. Lodge, "Space-time coding in mobile satellite communications using dual-polarized channels," *IEEE Transactions on Vehicular Technology*, vol. 55, no. 1, pp. 188–199, Jan 2006.

[13] "Satellite component of UMTS (S-UMTS); family SL satellite radio interface," *ETSI TS 102 744*, Oct. 2015.